\documentclass[aps,pre,superscriptaddress,showpacs,showkeys,twocolumn,floatfix]{revtex4}

\usepackage{graphicx}
\usepackage{dcolumn}
\usepackage{bm}
\usepackage{amsfonts}
\usepackage{amsmath}
\usepackage{amssymb}
\usepackage{latexsym}
\usepackage{color}
\usepackage[english]{babel}

\usepackage{textcomp}

\begin{document}
\def\myfrac#1#2{\frac{\displaystyle #1}{\displaystyle #2}}
\def\red#1{{\color{red} #1}}

\title{Characterization of azimuthal and radial velocity fields induced by rotors in flows with a low Reynolds number}

\author{J.~K\"{o}hler}
\affiliation{Applied Laser Technologies, 
Ruhr-Universit\"{a}t Bochum, Universit\"{a}tsstra\ss{}e 150, 44801 Bochum, Germany} 
\author{J.~Friedrich}
\affiliation{Insitut f\"{u}r Theoretische Physik I, Ruhr-Universit\"{a}t Bochum,
Universit\"{a}tsstra\ss{}e 150, 44801 Bochum, Germany}
\author{A.~Ostendorf}
\author{E.~L.~Gurevich}
\affiliation{Applied Laser Technologies, 
Ruhr-Universit\"{a}t Bochum, Universit\"{a}tsstra\ss{}e 150, 44801 Bochum, Germany} 

\date{\today}

\begin{abstract}
We theoretically and experimentally investigate the flow field that 
emerges from a rod-like microrotor rotating about its center in a non-axisymmetric manner.
A simple theoretical model is proposed that uses a superposition of two rotlets as a fundamental 
solution to the Stokes equation. 
The predictions of this model are compared to measurements of the azimuthal and 
radial microfluidic velocity field components that are induced by a rotor composed 
of fused microscopic spheres. The rotor is driven magnetically and the fluid flow is 
measured with the help of a probe particle fixed by an optical tweezer.
We find considerable deviations of the mere azimuthal flow pattern induced by a single rotating 
sphere as it has been reported by Di Leonardo \textit{et al.} [Phys. Rev. Lett. 96, 134502 (2006)].
Notably, the presence of a radial velocity component that manifests itself by an oscillation 
of the probe particle with twice the rotor frequency is observed.
These findings open up a way to discuss possible radial transport in microfluidic devices.
\end{abstract}

\maketitle

\section{Introduction}
A key aspect in microfluidics is to understand the theoretical fundamentals of flow at 
low-Reynolds numbers and to exploit them directly in applications and devices.
As the miniaturisation is an ongoing trend, especially lab-on-a-chip (LOC) or micro 
total analysis systems (\textmu TAS) provide a fast reaction 
time, low amount of reagents and low waste generation as well as high detection 
accuracy and verifiability \cite{Gijs2004}.
Therefore, the analysis and diagnostics of reagents within microfluidic channels require 
switching, driving and mixing of the liquid fluids \cite{Whitesides2006}.
Optical forces, for instance, provide a non-contact trapping and 
manipulation of microobjects or biological cells and are sufficient for usage 
in different microfluidic applications. The latter are 
e.g., pumping due to rotation of microstructures, 
sorting cells and intermixing different fluids \cite{Mohanty2012,Wu2011}.
Padgett \textit{et al.} identify the use of optical forces in so-called 
optical tweezers (OT) as a technology with high potential for 
microfluidic applications in biology, medicine or chemistry \cite{Padgett2011}.

Furthermore, the need of measuring systems within microfluidic channels 
with high precision arises. Besides actuation, systems based on optical 
forces can be used for sensing and characterizing of fluid properties.
For example, Di Leonardo \textit{et al.} and Kn\"{o}ner \textit{et al.}  
have measured the flow field in a microfluidic 
device with OT's \cite{Padgett2011,DiLeonardo2006,Knoener2005}. 
In both studies a spherical microrotor has been activated by interaction of polarized laser light 
with a birefringent vaterite crystal. The rotation has led to a fluid flow and 
a displacement of an inserted probe particle. The azimuthal flow field around the microsphere
has been determined by measuring the shift of the probe particle. 

However, no radial component has been detected in these studies. 
Objects in microfluidic systems rotating about their 
center in axisymmetric manner can easily produce azimuthal flow, which is also 
supported by theory \cite{Landau1966,Jeffery1922}. Whether a rotating microstructure can 
induce a radial flow component at low-Reynolds numbers, however, 
seems to be far less understood. Moreover, for axisymmetric rotations,
simple
symmetry arguments suggest that such a radial flow
can not occur.
In addition, theoretical works \cite{camassa,feng,viana} on 
non-axisymmmetric rotations appear to be restricted to ellipsoidal bodies and prove to be quite challenging from
a mathematical point of view. The work of Camassa \emph{et al.} \cite{camassa}, for instance,
treats the problem of an ellipsoid
sweeping out a double cone in Stokes flow. Here, a rather complex spatio-temporal velocity field
is obtained and explicitly analyzed for certain limit cases (small cone angles, far-field behavior etc.). Although
a radial flow component is predicted, the main focus of this works lies on the advection of tracers and
the establishment of certain invariants for the above mentioned limit cases.
Consequently, the main purpose of this paper is to gain some insight on the persistence of such 
radial flow for the case of non-axisymmetric rotations
at low-Reynolds numbers. Moreover, the azimuthal velocity field component is investigated and compared to
the case of a purely axisymmetric rotation. To this end, a comparatively 
simple model that is based on the superposition
of two rotlets \cite{blake,pozrikidis} is used in order to approximate the rod-shaped
rotor geometry and to make statements about the induced velocity fields.
 
The mere fact that the rotlet solution as a fundamental solution to the Stokes equation 
for low-Reynolds number flows basically shares the same velocity profile as the 
point-vortex solution of the two-dimensional Euler equation \cite{aref,saffman,chaos} allows to draw an 
interesting analogy to a recently proposed generalized vortex model \cite{Friedrich2013}.
Here, the model consists of pairs of like-signed point-vortices and it has been shown
that the straining of these small-scale vortical structures leads to non-vanishing radial 
velocity components that prove to be of great importance for the energy transport in 
such a two-dimensional flow. Although the rotlet and the point-vortex solution
apply to two very different Reynolds number regimes,
this striking analogy allows to predict a possible radial flow also
for the Stokes equation.

In this paper, the predictions of this rotlet model
are compared to experimental results that allow the detection of azimuthal and radial flow components
influenced by microrotors. 
The experiments were performed using a combination of optical and magnetic forces, which enables the 
manufacture of rod-shaped microrotors and the investigation of the flow fields.
Recently this method has been successfully applied for constructing a micro-\\ 
pump for microfluidic channels \cite{Koehler2014}.

\section{The rotlet model equations}
\begin{figure}[t!]
  \centering
   \includegraphics[width=0.45\textwidth]{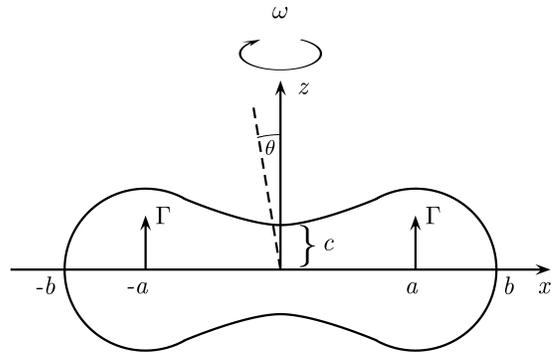}
   \caption{Approximation of a dumbbell-shaped body of length $b$ and neck length $c$
   by a pair of rotlets with strength $\Gamma$ fixed at $\pm {\bf a}$. 
   Initially,
   the body is supposed to be at rest as the surrounding fluid undergoes a rotation with the angular velocity
   $\omega$ determining the parameters $\Gamma$ and $a$ in terms of the rotor geometry $b$ and $c$.} 
   \label{rot_image}
   \end{figure}

\begin{figure}[t!]
  \centering
   \includegraphics[width=0.45\textwidth]{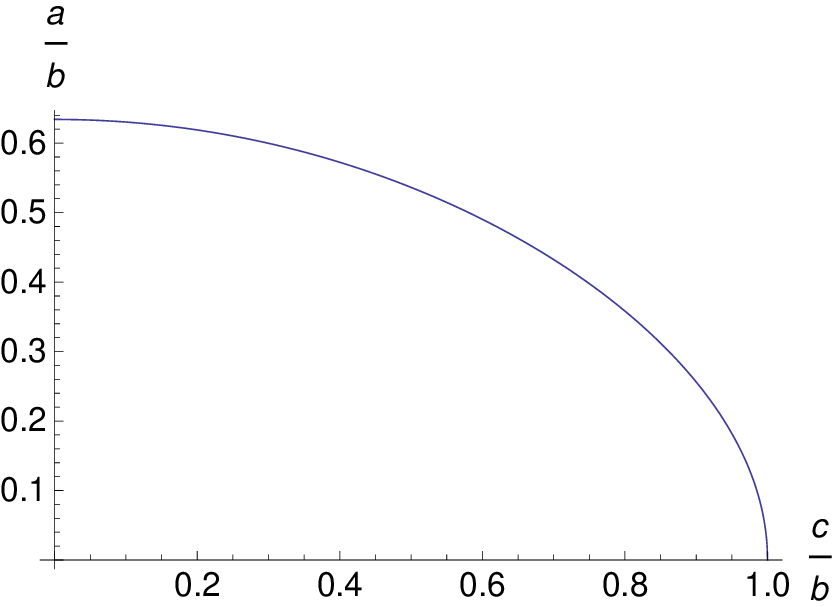}
   \caption{Non-dimensional parameter solution for ${a}/{b}$ of Eq. (\ref{root}) in dependence
   of ${c}/{b}$. 
   In the limit of ${c}/{b} \rightarrow 1$, the solution of a rotating sphere $a=0$ is recovered.} 
   \label{root_non_dimensional}
\end{figure}

The proposed model captures the main effects of
the rotors investigated in the experiment. Here, we mainly follow the procedure
described by Chwang and Wu who approximated the velocity field of a rotating dumbbell-shaped
body via a superposition of two rotlets spinning about the main axis of symmetry \cite{Chwang1974}; hence, 
the resulting velocity field only possessed an azimuthal component.
The main difference in our approach is that we consider non-axisymmetric rotations,
which turn out to induce an additional radial flow component. 
To this end, we initially work in a frame of reference where the rotor is at rest and assume
that the entire fluid is undergoing a rotation about the $z$ axis with the angular velocity 
$\omega$. We then consider the velocity field that emerges from two rotlets at 
${\bf a}$ and ${-\bf a}$
 \begin{equation}\label{rotlet-dipol}
  {\bf u}({\bf x})= \Gamma \omega  \left[ {\bf e}_z \times \frac{{\bf x}-{\bf a}}{|{\bf x}-{\bf a}|^3} 
		     + {\bf e}_z \times \frac{{\bf x}+{\bf a}}{|{\bf x}+{\bf a}|^3} \right] 
		     -  \omega {\bf e}_z \times {\bf x} 
 \end{equation}
 as an approximation of the velocity field that is induced by the rotor.
 Note that the last term has been added to ensure the correct fluid velocity in the far-field.
 Furthermore, the rotlet distance ${\bf a}$ and the rotlet strength $\Gamma$ have to be considered
 as free parameters yet to be determined by the rotor geometry.
 In order to see how these two paramters are fixed, we assume that the 
 dumbbell is oriented in ${\bf e}_x$-direction
 as it is depicted in Fig. \ref{rot_image}. 
 As a consequence, the velocity field in Eq. (\ref{rotlet-dipol}) has to satisfy no-slip
 boundary conditions at ${\bf x}_b= \pm b {\bf e}_x$ and
 at ${\bf x}_c= \pm c {\bf e}_y$, namely ${\bf u}({\bf x}_b) =0$ and
 ${\bf u}({\bf x}_c) = 0$.  
 From these boundary conditions, we obtain 
 \begin{equation}\label{boundary1}
  \frac{b^2 + a^2}{(b^2-a^2)^2}= \frac{b}{2 \Gamma} \qquad  (a^2+c^2)^{-3/2} = \frac{1}{2\Gamma}.
 \end{equation}
In eliminating $\Gamma$ from these equations, we obtain a relationship between 
the geometrical parameters 
\begin{equation}\label{root}
  \left(1 + \frac{a^2}{b^2}\right)\left(\frac{a^2}{b^2}+\frac{c^2}{b^2}\right)^{3/2}=
  \left(1-\frac{a^2}{b^2}\right)^2,
\end{equation}
which can be solved for a variety of rotor geometries in order to obtain the non-dimensional parameter
${a}/{b}$ in dependence of ${c}/{b}$. The corresponding solution for ${a}/{b}$ determining
the rotlet distance from the center is depicted in Fig. \ref{root_non_dimensional}.

At this point, it 
has to be stressed that the approximation (\ref{rotlet-dipol}) is not capable of reproducing the no-slip
boundary conditions over the entire rotor body but is only strictly valid at the local points
${\bf x}_b= \pm b {\bf e}_x$ and
 at ${\bf x}_c= \pm c {\bf e}_y$. 
However, in the following, we are not so concerned about the exact velocity field near the rotor
body but rather try to understand the immediate consequences of the velocity
field in Eq. (\ref{rotlet-dipol}) on the flow pattern in the far-field. 
To this end, we make use of the similarity of the rotlet velocity profile in Eq. (\ref{rotlet-dipol}) 
to the point-vortex solution of two-dimensional turbulence, which only differ by the power law in 
the denominator (quadratic instead of cubic for point-vortices). In this context, it has been 
shown recently that a generalized vortex model \cite{Friedrich2013} consisting of pairs of 
like signed point-vortices leads to a radial flow component in the far-field.

Applying these results to the rotlet pair of Eq. (\ref{rotlet-dipol}), a multi-pole
expansion for $|{\bf a}|/|{\bf x}| \ll 1 $ yields
\begin{equation}\label{multi}
 {\bf u}({\bf x})= \Gamma \omega \left[ 2 {\bf e}_z \times \frac{{\bf x}}{x^3} +
  {\bf e}_z \times( {\bf a} \cdot \nabla_{\bf x})^2  \frac{{\bf x}}{x^3} \right]-
  \omega {\bf e}_z \times {\bf x}. 
\end{equation}
\begin{figure*}
\begin{minipage}[l]{0.325 \textwidth}
\centering
\includegraphics[width=1 \textwidth]{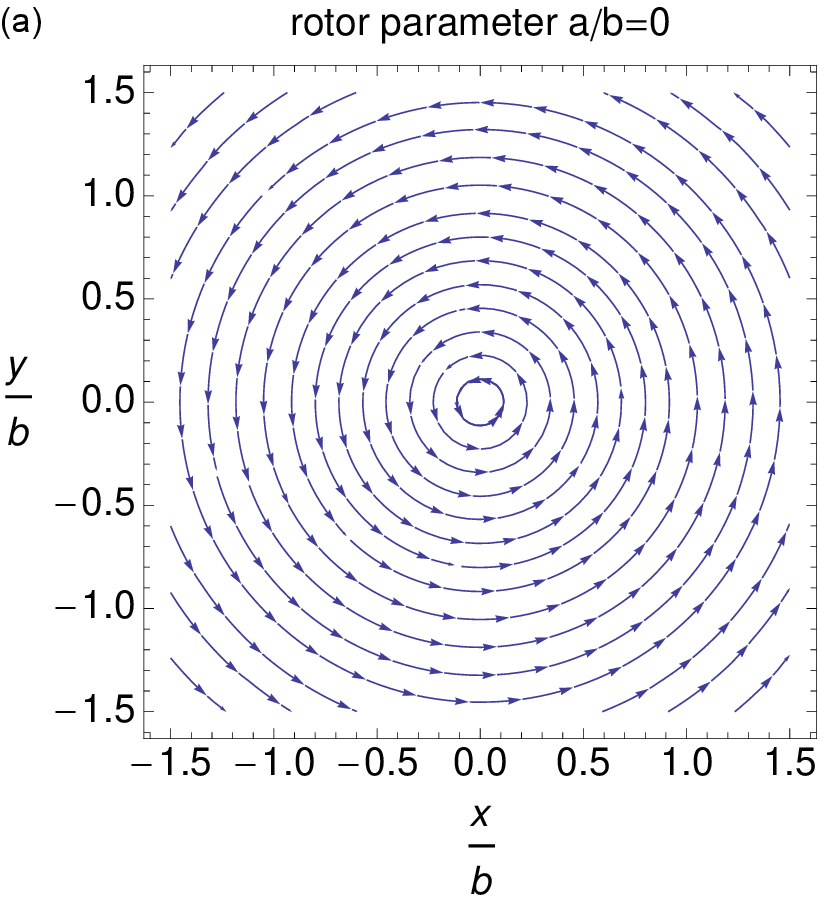}
\end{minipage}
\begin{minipage}[l]{0.325 \textwidth}
\centering
\includegraphics[width=1 \textwidth]{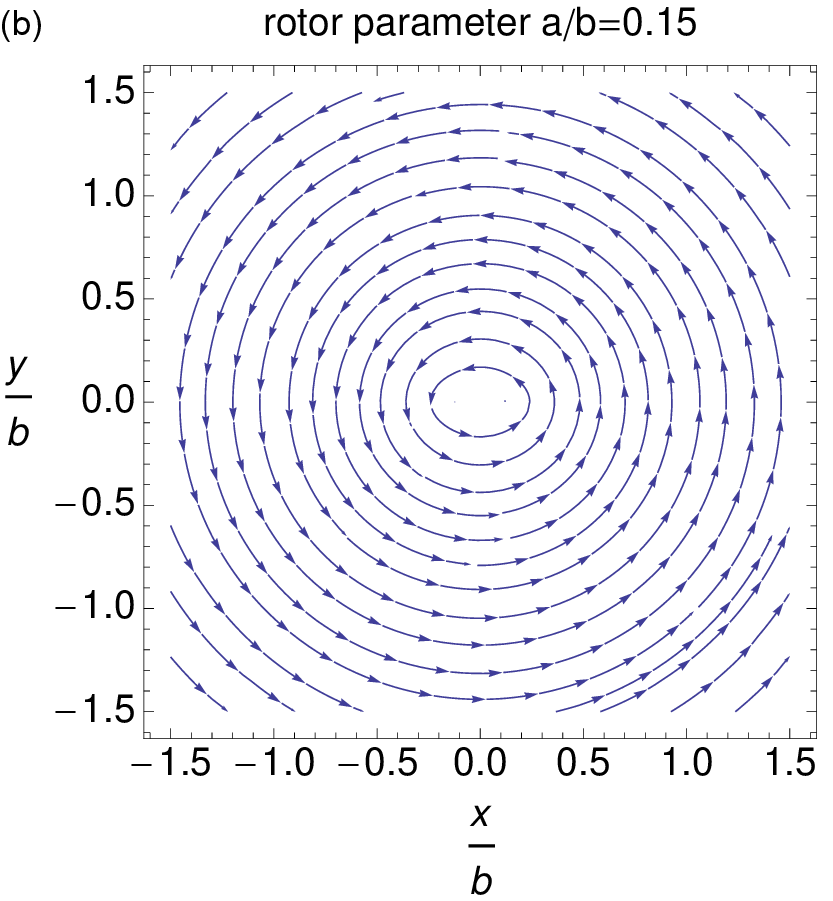}
\end{minipage}
\begin{minipage}[l]{0.325 \textwidth}
\centering
\includegraphics[width=1 \textwidth]{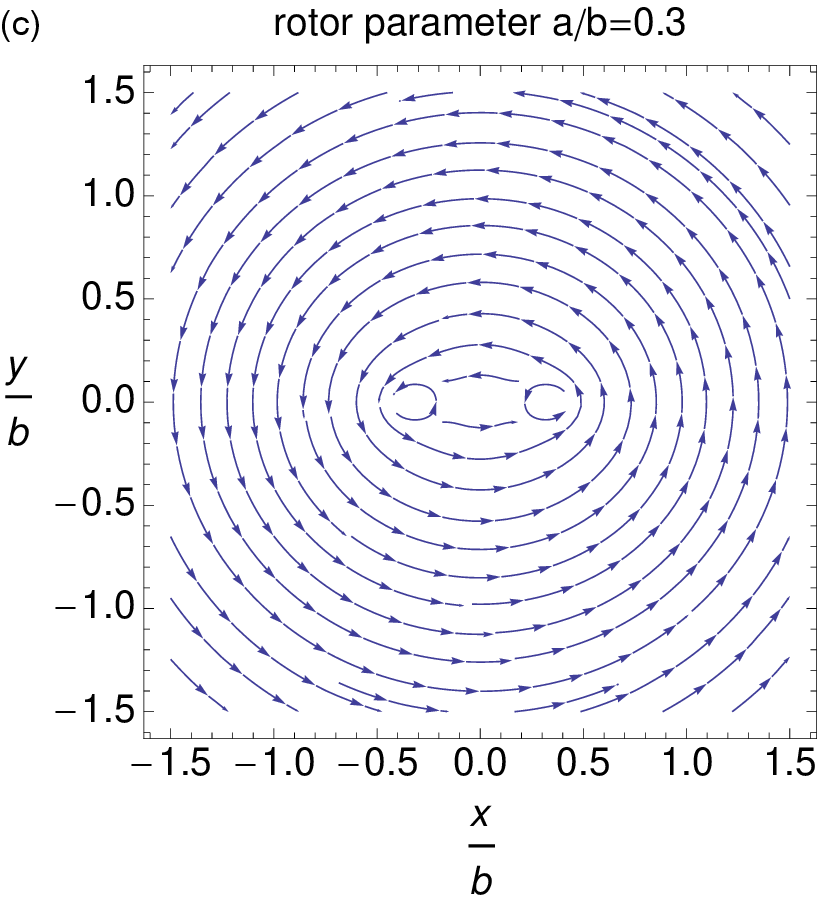}
\end{minipage}
\begin{minipage}[l]{0.1625 \textwidth}
\end{minipage}\\~
\\
~
\\
~
\\
\begin{minipage}[l]{0.325 \textwidth}
\centering
\includegraphics[width=1 \textwidth]{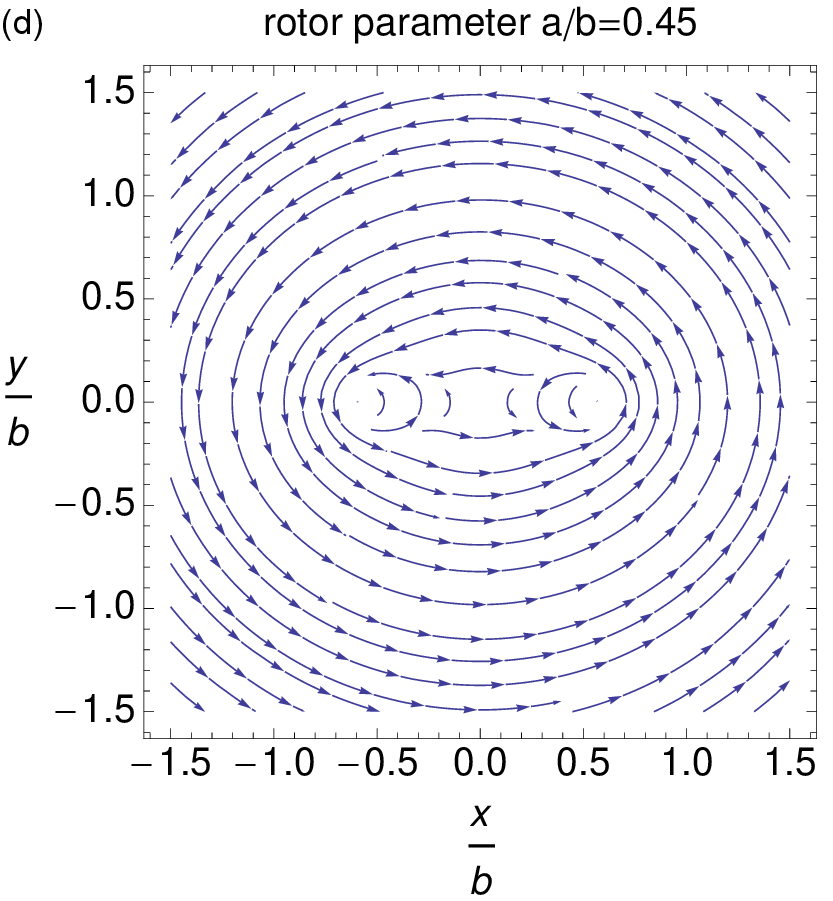}
\end{minipage}
\begin{minipage}[l]{0.325 \textwidth}
\centering
\includegraphics[width=1 \textwidth]{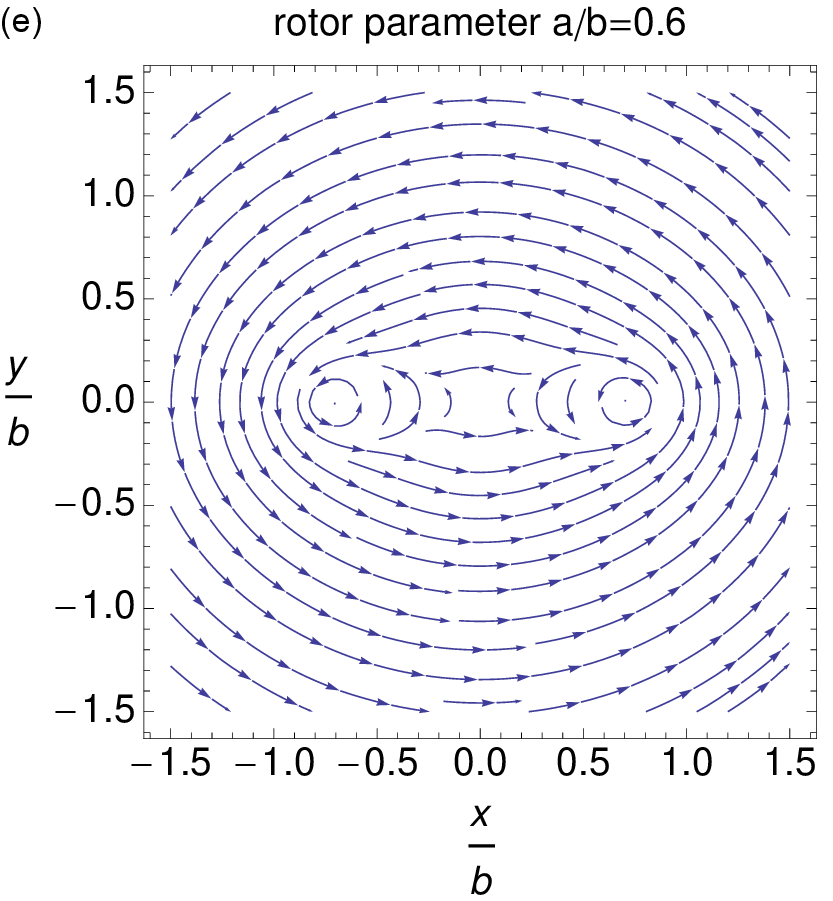}
\end{minipage}
\begin{minipage}[l]{0.1625 \textwidth}
\end{minipage}
\caption{Velocity field profiles in the $x$-$y$-plane determined by the far-field approximation (\ref{u_r}) and
(\ref{u_phi}) for increasing $a/b$ at $t=0$ (rotor is oriented in $x$-direction). The axisymmetric velocity field of the rotating sphere is recovered for $a=0$. As
$a/b$ approaches its maximum value determined by the solution of Eq. (\ref{root}) in Fig. \ref{root_non_dimensional} 
the velocity field becomes more and more distorted in the $x$-direction.}
\label{profile}
\end{figure*}
Here, the first term corresponds to the velocity field of a rotating sphere with radius 
$[2 \Gamma \omega]^{1/3}$, which solely leads to an azimuthal component of the velocity 
field.
The second term however, can lead to a component of the velocity in radial direction, which 
can be seen in writing the multi-pole terms in Eq. (\ref{multi}) explicitly as
\begin{equation}
 {\bf u}({\bf x})= \Gamma \omega {\bf e}_z \times  \left[ 2 \frac{{\bf x}}{x^3} 
 -6 \frac{{\bf a}}{x^5} {\bf a} \cdot {\bf x}  +15 \frac{{\bf x}}{x^7} ({\bf a} \cdot {\bf x})^2 -
 3 \frac{{\bf x}}{x^5} a^2 \right ]
  \end{equation}
The term that allows the velocity field to escape from its solely azimuthal shape is the second one, 
which can be seen in projecting 
onto the radial direction. To this end, we introduce the unit vector 
${\bf e}_a=\frac{{\bf
a}}{a}=\left( \begin{array}{l}     
    \cos \varphi_a  \\
    \sin \varphi_a \\ ~~~0
    \end{array}
    \right)$, as well as
 ${\bf e}_r=\frac{{\bf x}}{x}=\left( \begin{array}{l}     
    \cos \varphi \sin \theta  \\
    \sin \varphi\sin \theta \\
    ~~~\cos \theta
    \end{array} \right)$, which yields
\begin{equation}
{\bf e}_r \cdot 
[{\bf e}_z \times {\bf e}_a]({\bf e}_a\cdot {\bf e}_r)=
\frac{1}{2} \sin(2 (\varphi -\varphi_a)) \sin^2 \theta.
\end{equation} 
Additionally, we change to a frame of reference in which the fluid is at rest and 
the rotor is undergoing rotation, i.e.
$\varphi - \varphi_a \rightarrow \varphi -  \omega t$ 
and forget about additional centrifugal forces due to the neglection of inertia
in the Stokes equation ( $\rho | \boldsymbol \omega \times {\bf u}|$ small compared to
the viscous force $\mu |\nabla^2 {\bf u}|$, 
where $\rho$ is the density and $\mu$ is the viscosity of the fluid). 
We obtain
\begin{eqnarray} \label{u_r}
 u_r  ={\bf e}_{r} \cdot {\bf u}({\bf x})
= -3 \Gamma \omega \frac{a^2}{r^4} \sin(2 ( \varphi-\omega t )) \sin^2 \theta 
\end{eqnarray}
for the radial velocity component and 
\begin{eqnarray}  \label{u_phi}
 &u_{\varphi}& = {\bf e}_{\varphi} \cdot {\bf u}({\bf x}) \\ \nonumber
 &=& \Gamma \omega \left( \frac{2}{r^2}\sin \theta 
 - 3 \frac{a^2}{r^4}\sin \theta
 +15 \frac{a^2}{r^4} \sin^2 \theta \sin^2 ( \varphi-\omega t) \right)
\end{eqnarray}
for the azimuthal velocity component. It can be seen that the radial component oscillates 
with twice the rotor frequency, whereas the azimuthal component is a superposition of
a time-independent part and a time-dependent part that does not vanish in average, since
$\langle  \sin^2 ( \varphi-\omega t) \rangle$ is a constant. Thereby the time-dependence
of the velocity field not inherently contained in the Stokes equation is due to the inclusion
of time-dependent boundary conditions similarly to the case of Taylor's swimming sheet \cite{taylor}.
The corresponding velocity field determined by the radial (\ref{u_r}) and azimuthal (\ref{u_phi}) flow components 
for increasing ${a}/{b}$ are shown in Fig. \ref{profile} for $t=0$.
It can be seen that the axisymmetric velocity field of the sphere ${a}/{b}=0$ gets distorted in the $x$-direction
for increasing ${a}/{b}$ as the two rotlets in Eq. (\ref{rotlet-dipol}) become more and more remote.

\section{Assembling, actuation and measurements}
\begin{figure}[t!]
\centering
  \includegraphics[width=0.45\textwidth]{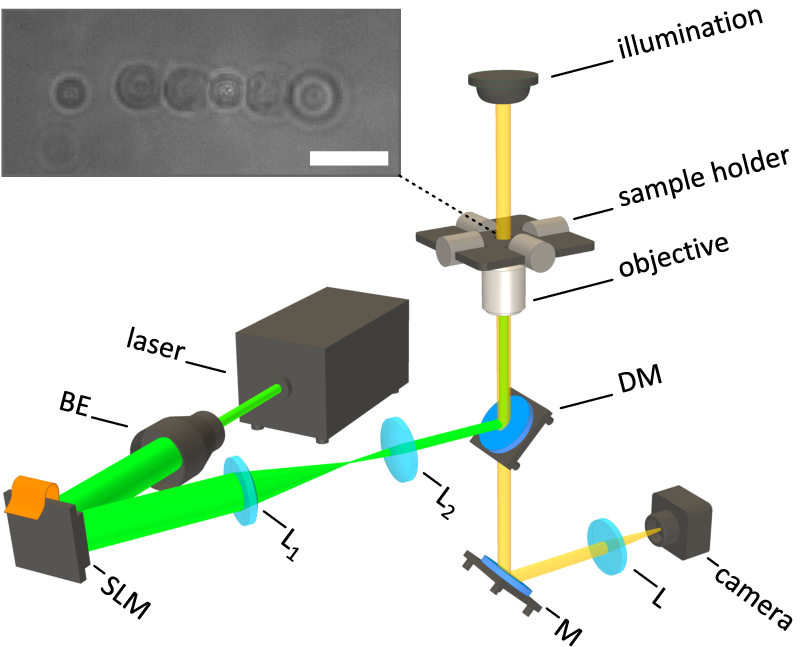}
  \caption{(color online) Schematic depiction of the
holographic optical tweezer (HOT) setup. The laser beam is modulated
and reflected by the spatial light modulator (SLM) and focused with a high-NA microscope objective into the sample chamber. 
The high-speed camera is used to observe and detect the particles in the
trapping plane. BE: beam expander, L: lens, M: mirror, DM: dichroic mirror. 
Inset: Microscope image of probe particle next to assembled five-particle rotor (scale bar is 5\,\textmu m).}
  \label{laser_setup}
\end{figure}
To verify the established theoretical model a microrotor has to be assembled, actuated and the generated fluid flow has to be measured experimentally in a precise manner.
The OT is used to assemble chain-shaped magnetic microrotors by a simple ``pick-and-place'' technique \cite{Ghadiri2012}.
Based on a focused laser beam, OT provide non-contact trapping, moving and rotation of dielectric microparticles \cite{Ashkin.1986}. The force acting on a particle is typically in the range of several piconewtons and can be described by
\begin{equation}
  {\bf F}_{opt} = -k \mathfrak{r}
 \label{hook}
\end{equation}
with the trap stiffness $k$ and the displacement of the particle ${\bf \mathfrak{r}}$ from its equilibrium position.
A Biotin (B) coated transparent microsphere (SiO2-Biotin-AR722, $\varnothing$ = 2.73 $\pm$ 0.13 \textmu m, microParticles GmbH) is trapped by OT and fused together with Streptavidin (SA) coated magnetic microspheres (SiO2-MAG-SA-S2548, $\O$ = 2.61 $\pm$ 0.12 \textmu m, microParticles GmbH) by bringing them into contact. The mutual high affinity of the biomolecules leads to a stable binding of the complementary particles and a magnetic three-particle rotor can be assembled \cite{Ghadiri2012}.
The embedded transparent microsphere in the middle of the rotor provides stable trapping and friction-free rotation about the rotational axis of the rotor.
For bigger rotors magnetic interaction is used to attach magnetic particles at both ends of the microstructure (Fig. \ref{laser_setup} inset). Although, the outer magnetic particles of the five-particle rotor are not connected with the B-SA binding, there is no deformation of the rotor up to the rotational frequency of 6.5\,s$^{-1}$.
The following experimental investigation is based on chain-shaped rotors with three and five particles.
\begin{figure}[t!]
\centering
  \includegraphics[width=0.45\textwidth]{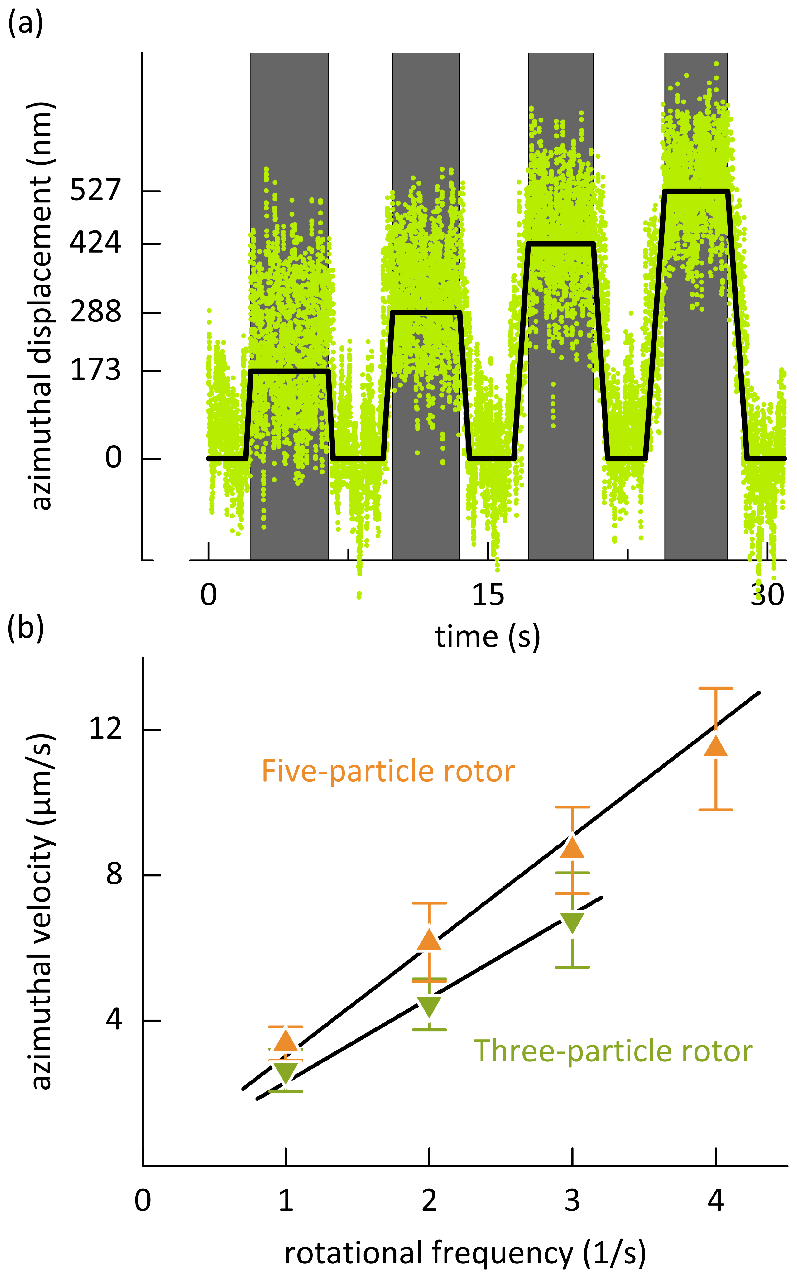}
  \caption{(color online) a) The displacement of the probe particle increases with higher rotational frequency of the five-particle rotor. b) The calculated fluid velocities of five- and three-particle rotors show linear behavior.} 
  \label{a_flow1}
\end{figure}

A cw laser (Verdi G5, Coherent) with a wavelength of $\lambda = 532$\,nm is used and with integration of a spatial light modulator (SLM) multiple optical trapping spots can be generated (Fig. \ref{laser_setup}). Computer generated holograms implemented into the SLM allow simultaneous manipulation of multiple microobjects using a single laser source. The SLM based OT's are commonly referred to as holographic optical tweezers (HOT) \cite{Hayasaki1999,Curtis2002,Grier2006}. 
The laser beam is expanded to utilize the entire active area of the SLM (Pluto-VIS, Holoeye) and the reflected modulated beam is projected onto the back aperture of a 100\,x 1.4\,NA microscope objective using two lenses (L$_1$ and L$_2$) with 4-f-configuration. The beam is focused into the sample chamber and the microparticles can be trapped in an aqueous environment. The sample chamber has a circular aperture
of 8\,mm diameter, a height of 5\,mm and is closed on the top and bottom with cover
glasses. A high-speed CMOS camera (USB UI-1225LE, IDS) is used to observe the particle position and to analyze it with sub-pixel accuracy down to 5\,nm \cite{Neuman2004}.
The rotation of the assembled microstructures is introduced by a controllable external magnetic field, which is generated by four electromagnetic coils arranged around the sample chamber and addressed by a stepper motor driver (TMCM-1110, Trinamic). Due to the magnetic torque, the microrotor follows the external magnetic field and rotates with desired rotational speed.
\begin{figure}[t!]
\centering
  \includegraphics[width=0.45\textwidth]{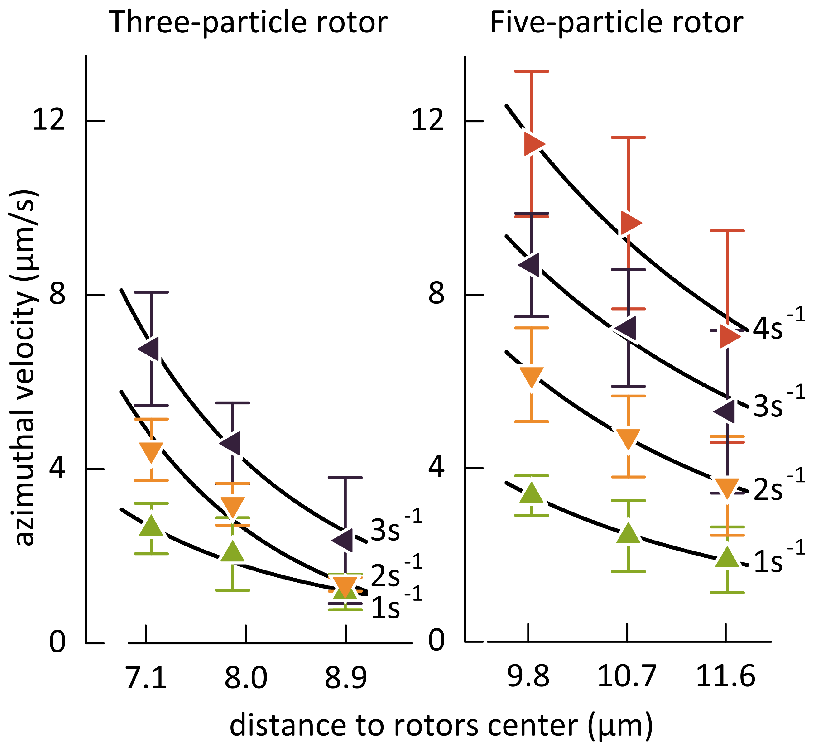}
  \caption{(color online) The azimuthal fluid velocity decreases with increasing distance to the rotor. 
  The results are in good agreement to established Eq. \eqref{u_phi} which is illustrated by the fits (black lines).}
  \label{a_distance1}
\end{figure}
\begin{figure}[h!]
\centering
  \includegraphics[width=0.45\textwidth]{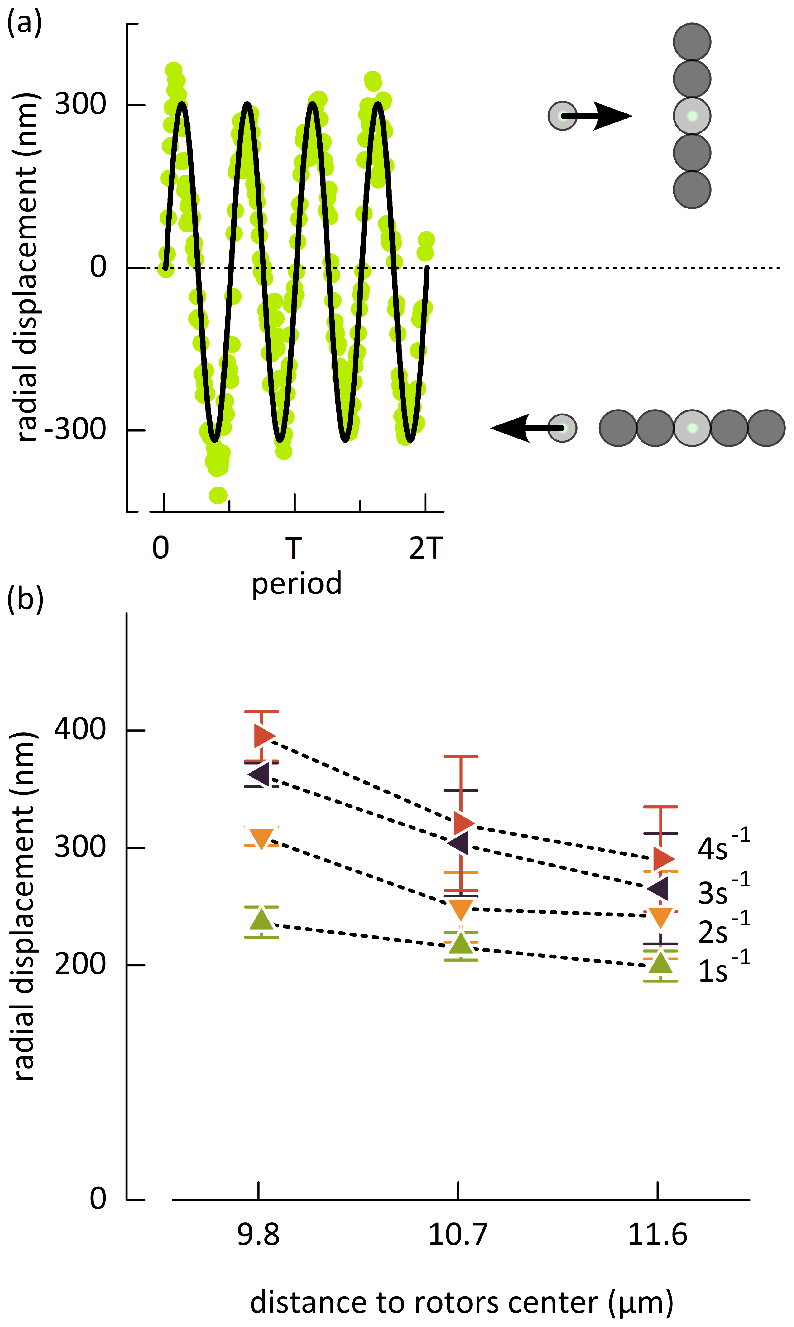}
  \caption{(color online) a) The oscillating radial displacement of the probe particle (green dots) is presented at a rotational frequency of 4\,s$^{-1}$. The black curve represents a sine function fit which reaches its maximum (minimum) when the rotor is vertically  (horizontally) aligned, respectively. b) The radial displacement of the probe particle decreases with increasing the distance to the rotor.
  The dashed lines connect the corresponding points for the sake of clarity.}
  \label{r_displacement}
\end{figure}
The rotation leads to a flow field in the liquid environment, which is measured by introducing a non-magnetic probe particle as illustrated in the inset of Fig. \ref{laser_setup}. The probe particle is trapped by a second optical spot at a variable distance to the rotor axis. 
The generated fluid flow ${\bf u}$ leads to a displacement of the probe particle which can be calculated using Stroke's law (drag force),
\begin{equation}
  {\bf F}_{drag} = 6 \pi \eta R_p   c_F {\bf u}
\end{equation}
with the fluid viscosity $\eta$ (the experiments are performed in buffer solution contains approx. 99\% water), the probe particle radius $R_p$ and the correction factor $c_F$. Due to boundary effects of the surfaces the correction coefficient has to be implemented according to Faxen's law \cite{Svoboda1994}, which is calculated to be $c_F$ = 1.62 for these experiments. As the probe particle will remain in the equilibrium position defined by the optical force and the drag force, the fluid velocity can than be expressed by
\begin{equation}
  {\bf u} = \frac{k\, \mathfrak{r}}{6\, \pi\, \mu\, R_p \, c_F}.
  \label{flow_eq}
\end{equation}
The equipartition theorem \cite{Neuman2004} is used in order to  evaluate the trap stiffness $k$.

\section{Results}
Fig. \ref{a_flow1} a) shows the displacement of the probe particle in azimuthal direction by varying the rotational frequency from 1\,$^{-1}$ to 4\,s$^{-1}$ of a five-particle rotor. The rotational magnetic field is applied for approx. 4\,s for each measurement followed by a pause period to clearly identify the distinct regions. Here, the rotor radius is $R_r$ = 7.2\,\textmu m and the distance of the probe particle to the rotor center is $r$ = 9.8\,\textmu m. Note that the measured shift of the probe particle is constant for a steady rotation (gray shaded areas) and is independent on the instantaneous rotor orientation. That indicates that the last term of Eq. \eqref{u_phi} is averaged during the measurements.
As depicted in Fig. \ref{a_flow1} b) the azimuthal velocity linearly increases with the rotational frequency.
The averaged azimuthal velocities are shown for the five-particle rotor corresponding to Fig. \ref{a_flow1} a) and a three-particle rotor with $R_r$ = 4.4\,\textmu m at the distance of $r$ = 7.1\,\textmu m at different rotational frequencies, exemplary.
In evidence, the five-particle rotor generates higher fluid flow as the three-particle rotor.

In Fig. \ref{a_distance1} the azimuthal fluid velocity in dependence of the distance to the rotor is illustrated for the two different rotor types. The measured points represent the average values of five measurements  at constant conditions.
The graphs reveal that there is a marked decrease of the fluid velocity with increasing the distance to the rotor. The plotted black lines correspond to fits using Eq. \eqref{u_phi} namely $u_{\varphi} = \alpha r^{-2} + \beta r^{-4}$, whereas in \cite{DiLeonardo2006,Landau1966} for a rotating sphere only decrease as $r^{-2}$ has been reported.

Eq. \eqref{u_r} indicates that the radial velocity component leads to an oscillating fluid flow with twice the rotor frequency. Fig. \ref{r_displacement} a) highlights this characteristic by showing the radial displacement of the probe particle relative to the period $T = 2 \pi / \omega$.
The oscillation is measured by a rotor with $R_r = 7.2$\,\textmu m at the distance $r = 9.8$\,\textmu m and rotational frequency of 4\,s$^{-1}$.
The black curve represents a sine function fit.
The period calculated with discrete Fourier transform (DFT) is the half of the period of the external magnetic field and thus confirms to the theoretical model.
The maximum positive displacement of the probe particle is reached when the rotor is vertical aligned. Vice versa the probe particle experiences its maximum negative displacement when the rotor is horizontal aligned.

At this point it is not possible to calculate the exact value of the radial fluid velocity. The constant shift of the probe particle in azimuthal direction (cf. Fig. \ref{a_flow1} a)) leads to an unknown change of optical trapping force. Hence, Eq. \eqref{flow_eq} can not be applied here. However, it can be noted that higher rotating frequencies lead to an increased amplitude of the displacement (Fig. \ref{r_displacement} b)).

\section{Conclusion}
In conclusion, we have presented theoretical and experimental evidence that a non-axisymmetric rotation
of a rod-like rotor at low-Reynolds number flow leads to a quite complex
spatio-temporal flow pattern.
The validity of the established theoretical model is demonstrated by means of experimental investigations for azimuthal and radial flow fields.
In particular, the presence of an oscillating radial velocity component not generated in purely axisymmetric rotations \cite{Landau1966} is a quite intriguing new feature.
It will be a task for future experiments to further classify the radial flow component especially with respect to its power law behavior with increasing distance from the rotor.
These findings might be important for mixing and driving 
in microfluidic devices as well as for the theory of 
swimming at low-Reynolds numbers.
Furthermore, this experimental technique provides the measurement of fluid flow in microfluidic systems with high precision.

\section{Ackowledgments}
J.K. and A.O. are greatful for support within the Reinhardt Koeselleck project (OS 188/28-1) of the German Research Foundation 
(DFG).






\begin{thebibliography}{10}

\bibitem{Gijs2004}
M.~A.~M. Gijs,
\newblock Microfluid. Nanofluid. {\bf 1} (2004).

\bibitem{Whitesides2006}
G.~M. Whitesides,
\newblock Nature {\bf 442}, 368 (2006).

\bibitem{Mohanty2012}
S.~Mohanty,
\newblock Lab Chip {\bf 12}, 3624 (2012).

\bibitem{Wu2011}
J.~Wu and M.~Gu,
\newblock J. Biomed. Opt. {\bf 16}, 080901 (2011).

\bibitem{Padgett2011}
M.~Padgett and R.~Di~Leonardo,
\newblock Lab Chip {\bf 11}, 1196 (2011).

\bibitem{DiLeonardo2006}
R.~Di~Leonardo et~al.,
\newblock Phys. Rev. Lett. {\bf 96}, 134502 (2006).

\bibitem{Knoener2005}
G.~Kn\"oner, S.~Parkin, N.~R. Heckenberg, and H.~Rubinsztein-Dunlop,
\newblock Phys. Rev. E {\bf 72}, 031507 (2005).

\bibitem{Landau1966}
L.~Landau and E.~Lifshitz,
\newblock {\em Course of Theoretical Physics. Vol. 6: Fluid Mechanics},
\newblock London, 1966.

\bibitem{Jeffery1922}
G.~Jeffery,
\newblock Proc. R. Soc. A {\bf 102}, pp. 161 (1922).

\bibitem{camassa}
R.~Camassa, T.~J.Leiterman, and R.~M. McLaughlin,
\newblock J. Fluid Mech. {\bf 612}, 153 (2008).

\bibitem{feng}
J. Feng et~al.,
\newblock J. Fluid Mech. {\bf 283}, 1 (1995). 

\bibitem{viana}
F. Viana et~al.,
\newblock J. Appl. Math.{\bf 2005}, 341 (2005).

\bibitem{blake}
J.~Blake and A.~Chwang,
\newblock J. Eng. Math. {\bf 8}, 23 (1974).

\bibitem{pozrikidis}
C.~Pozrikidis,
\newblock {\em Boundary Integral and Singularity Methods for Linearized Viscous
  Flow},
\newblock Cambridge University Press, 1992,
\newblock Cambridge Books Online.

\bibitem{aref}
H.~Aref,
\newblock Ann. Rev. Fluid Mech. {\bf 15}, 345 (1983).

\bibitem{saffman}
P.~Saffman,
\newblock {\em Vortex Dynamics},
\newblock Cambridge Monographs on Mechanics, Cambridge University Press, 1992.

\bibitem{chaos}
J.~Argyris et~al.,
\newblock {\em An Exploration of Dynamical Systems and Chaos: Completely
  Revised and Enlarged Second Edition},
\newblock Springer Berlin Heidelberg, 2015.

\bibitem{Friedrich2013}
J.~Friedrich and R.~Friedrich,
\newblock Phys. Rev. E {\bf 88}, 053017 (2013).

\bibitem{Koehler2014}
J.~K\"{o}hler et~al.,
\newblock J. Phys. D: Appl. Phys. {\bf 47}, 505501 (2014).

\bibitem{Chwang1974}
A.~T. Chwang and T.~Wu,
\newblock J. Fluid Mech. {\bf 63}, 607 (1974).

\bibitem{taylor}
G.~Taylor,
\newblock Proc. R. Soc. A {\bf 209}, 447 (1951).

\bibitem{Ghadiri2012}
R.~Ghadiri, T.~Weigel, C.~Esen, and A.~Ostendorf,
\newblock J. Micromech. Microeng. {\bf 22}, 065016 (2012).

\bibitem{Ashkin.1986}
A.~Ashkin et~al.,
\newblock Opt. Lett. {\bf 5}, 11 (1986).

\bibitem{Hayasaki1999}
Y.~Hayasaki, M.~Itoh, T.~Yatagai, and N.~Nishida,
\newblock Opt. Rev. {\bf 6}, 24 (1999).

\bibitem{Curtis2002}
J.~E. Curtis, B.~A. Koss, and D.~G. Grier,
\newblock Opt. Commun. {\bf 207}, 169 (2002).

\bibitem{Grier2006}
D.~G. Grier and Y.~Roichman,
\newblock Appl. Opt. {\bf 45}, 880 (2006).

\bibitem{Neuman2004}
K.~C. Neuman and S.~M. Block,
\newblock Rev. Sci. Instrum. {\bf 75}, 2787 (2004).

\bibitem{Svoboda1994}
K.~Svoboda and S.~M. Block,
\newblock Ann. Rev. Biophys. Biomol. Struct. {\bf 23}, 247 (1994).

\end{thebibliography}

\end{document}